\documentclass[fleqn,preprint,showpacs,preprintnumbers,amsmath,amssymb]{revtex4}
\usepackage{graphicx}
\usepackage{dcolumn}
\usepackage{bm}

\begin{document}

\title{Exact Nonlinear Excitations in Double-Degenerate Plasmas}
\author{M. Akbari-Moghanjoughi}
\affiliation{Azarbaijan University of
Tarbiat Moallem, Faculty of Sciences,
Department of Physics, 51745-406, Tabriz, Iran}

\date{\today}
\begin{abstract}
\textbf{In this work we use the conventional hydrodynamics (HD) formalism and incorporate the Chew-Goldberger-Low (CGL) double-adiabatic theory to evaluate the nonlinear electrostatic ion excitations in double-degenerate (electron spin-orbit degenerate) magnetized quantum plasmas. Based on the Sagdeev pseudopotential method an exact general pseudopotential is calculated which leads to the allowed Mach-number range criteria for such localized density structures in an anisotropic magnetized plasma. We employ the criteria on the Mach-number range for diverse magnetized quantums plasma with different equations of state (EoS). It is remarked that various plasma fractional parameters such as the system dimensionality, ion-temperature, relativistic-degeneracy, Zeeman-energy, and plasma composition are involved in the stability of an obliquely propagating nonlinear ion-acoustic wave in a double-degenerate quantum plasma. Current study is most appropriate for nonlinear wave analysis in the dense astrophysical magnetized plasma environments such as white-dwarfs and neutron-star crusts where the strong magnetic fields can be present.}
\end{abstract}


\maketitle

\section{Introduction}

\textbf{Ion acoustic waves are of important tools to probe the plasma instabilities and responses to external perturbations. Two different methods for investigation of nonlinear behavior of ion acoustic waves are the reductive perturbation \cite{davidson} and the pseudopotential \cite{vedenov, sagdeev} methods. The reductive perturbation is used to evaluate the ion excitations in plasma with small- but finite-amplitude perturbation, while, the pseudopotential approach concerns the arbitrary-amplitude plasma excitation. The experimental realization of ion-acoustic waves dates back to 1970 \cite{ikezi}. There have been extensive past studies concerning the role of magnetism on oblique ion-acoustic wave propagations \cite{lee, witt, yashvir, yadav} in a classical plasma. Some later works extend the previous studies to the nonlinear wave dynamics to plasmas with ion-temperature anisotropy in magnetized plasmas \cite{cheong, mahmood}. The later works incorporate the Chew-Goldberger-Low (CGL) \cite{chew} double-adiabatic HD-model to evaluate the propagation of nonlinear ion-waves in dusty and electron-ion plasmas, respectively. In a sufficiently collisionless plasma the parallel and perpendicular ion-temperature may differ leading to double adiabaticity. Canuto and Chuo have investigated the collisionless single-fluid electron plasma dispersion in the presence of strong magnetic field in different electron number-density regimes \cite{can0} and have found that in the degenerate plasma regime the parallel and perpendicular electromagnetic propagations differ substantially. Unlike the classical plasmas, for a degenerate plasma in the zero-temperature limit, the anisotropy is not introduced due to the temperature anisotropy, but, is caused mainly because of density anisotropy which is related to the so-called Fermi-temperature of the degenerate system. In a degenerate plasma it is known that the electron collisions are much reduced because of Pauli-blocking mechanism. However, one should not confuse the Fermi-temperature of degenerate plasma with the real physical temperature of species. Such confusion has been shown to lead to converse results \cite{akbari00}. More recently, Shukla and Stenflo have introduced the CGL formulation to the quantum Hall magnetohydrodynamics model \cite{stan}.}

Since the early investigations by Bohm, Pines and Levine et.al \cite{bohm, pines, levine} major investigations have been directed towards the study of physical properties of dense plasmas the so-called quantum plasmas. Due to the significance of Pauli exclusion mechanism in such plasmas rather distinct features such as quantum tunneling, degeneracy pressure, quantization of electronic density of states and many other peculiar features \cite{landau, sho} are expected to occur which are absent in the ordinary plasma kind. Such quantum effects can even lead to distinct collective phenomena in quantum plasmas. Recent studies present some interesting features of the quantum plasmas \cite{haas1, haas2, Markowich, Marklund1, Brodin1, Marklund2, Brodin2, manfredi, shukla, gardner} unobserved in classical counterparts. Despite numerous applications of quantum plasmas in areas such as semiconductors, nanotechnology, quantum optics and electronics \cite{haug}, there are many astrophysical occasions where quantum effects become dominant. Chandrasekhar \cite{chandra1, chandra2, chandra3} has shown that a compact stellar object such as white-dwarf or neutron star can be effectively modeled as an ideal zero-temperature Fermi plasmas. It is also well-understood that the hydrostatic equilibrium of white-dwarfs or neutron stars is due to the electron or neutron degeneracy pressures in a degenerate superdense electron or neutron gas pressurized under the large stellar gravity force. The gravitational collapse may eventually setup in a super heavy star because of the softening of degeneracy pressure of relativistically moving fermions. The pressure loss may even lead to the ultimate collapse of the compact star under its own gravity. In such relativistically degenerate quantum plasmas the whole thermodynamical properties of the matter has been shown to alter \cite{kothary} and distinct features emerge in nonlinear wave dynamics.

On the other hand, interesting features emerge when the magnetic field and consequently the Landau orbital-quantization takes place in degenerate plasmas. Perhaps the most extensive theoretical investigation of thermodynamical properties of matter under arbitrary strength magnetic field belongs to that of Canuto et al. \cite{can1, can2, can3, can4}. These studies showed that, the thermodynamics and magnetic properties of a relativistically degenerate Fermi-Dirac gas is altered substantially due to spin-orbit degeneracy in the transverse direction to the field and differ fundamentally regarding to those of a normally degenerate ones. The magnetic susceptibility has been shown to exhibit the oscillatory behavior and the plasma may even become one-dimensional for very sufficiently high magnetic field strength. In such regime the fermion density is ruled strongly by both the magnetic field and relativity parameters. It was also found that when the plasma degeneracy becomes relativistic a self contained metastable ferromagnetic state may develop leading to strong magnetic field. This has been attributed to the presence of strong fields estimated for some highly magnetic compact stars \cite{con1, con2, con3, burk1, burk2}. More recently, quantum magnetohydrodynamic (QMHD) model has been extended to include the spin-1/2 effects \cite{Marklund3}. Investigations \cite{Marklund4, misra1, martin, mushtaq, brodin3, zaman, vitaly, misra2} based on spin-1/2 MHD model indicate that a negative pressure-like term in the momentum equation appears due to the electron spin-1/2 effects and dominates at very low temperatures leading to detectable features even in the presence of relatively small laboratory-scale magnetic field strengths.

Theoretical calculations reveal that strong kinetic-pressure anisotropy can be induced in the plasma due to the application of strong magnetic fields of order $4.414\times 10^{13}G$ which is typical of neutron stars or some white dwarf star cores. It has been shown that \cite{akbari0}, the mechanism known as Landau orbital ferromagnetism, usually referred to as LOFER is consistent with the large magnetic fields present in compact astrophysical objects \cite{crut, kemp, put, jor} and may in fact play a key role in a transverse instability such as transverse magnetic collapse or quantum collapse. Extensive reviews on the properties of matter under arbitrary strength magnetic field can be found in some recent literature \cite{dong, harding}. Other recent investigations \cite{chai, akbari7} also confirm the possibility of a lateral magnetic collapse in the relativistically degenerate magnetized plasma indicating that the highly magnetized gas can in fact experience a strong anisotropic stress due to the Lorentz-force and plasma magnetization. The later forces may completely cancel that of the electron degeneracy causing a magnetic collapse in equatorial plane. It has also been revealed that, the magnetism \cite{suh} and interaction effects such as Coulomb \cite{salpeter} can also slightly alter the Chandrasekhar mass-radius ratio. Distinct regimes have been shown to exist in a dense highly magnetized plasma due to Landau quantization \cite{mend1, mend2}. In current investigation we take into account the effect of anisotropy caused by uniform magnetic field on electronic-distribution on ion-acoustic wave propagation which is absent in the classical Chew-Goldberger-Low (CGL) double-adiabatic treatment. As will be discussed in later sections the effect of the electron-distribution anisotropy can become very important for the case of degenerated magnetized plasmas and may lead to instability of a highly spin-orbit quantized degenerate plasma. We use the conventional quantum hydrodynamic (QHD) model in Sec. \ref{equations} to extend the Chew-Goldberger-Low (CGL) double-adiabatic theory to include the electron pressure anisotropy in degenerate magnetized quantum plasmas and exactly solve for the solitary density excitations using conventional Sagdeev pseudopotential method in Sec. \ref{Sagdeev}. In later sections we give extensive examples of quantum plasmas for which the theory can be applied and draw our conclusions in Sec. \ref{conclusion}.

\section{Double-Degenerate Hydrodynamic Formulation}\label{equations}

In a uniformly magnetized quasineutral Fermi-Dirac plasma with arbitrarily high magnetic field strength is has been shown that the kinetic stress-tensor is highly anisotropic so that the transverse electron pressure due to sum of spin-orbit and degeneracy totally vanishes at the ground state Landau spin-orbit magnetization level \cite{can2}. Although, it has been claimed \cite{dong} that this anisotropy vanishes due to the work required to compensate the Lorentz magnetic-force acting on electrons, however, it can be shown that at the quantum-limit ($l=0$) of a Fermi-Dirac magnetized plasma the total pressure caused by the sum of the electron degeneracy pressure, $P_{e\parallel}$, and the spin-orbit magnetization pressure, $B\Gamma_e$, ($\Gamma_e$ is the plasma electron spin and orbital magnetization) still vanishes completely. In fact, the parallel and perpendicular electron pressure components in a uniformly magnetized plasma are related through the equation $P_{e\parallel}=P_{e\perp}+B\Gamma_e$ \cite{dong}. Using the definitions $\rho_e(r,B) \propto \partial {P_{e\parallel}{(r,B)}}/\partial {\epsilon_{Fe}}$ ($\epsilon_{Fe}=\sqrt{1+r^2}$ is the relativistic Fermi-energy of electrons) and $\Gamma_e(r,B) \propto {{\partial {P_{e\parallel}}({r},B)}}/{{\partial B}}$, with $r$ being the fractional relativistic Fermi-momentum. Thus, for a homogenous magnetized plasma, it is observed that only for the classical-limit ($l=\infty$), where the Chandrasekhar equation of state, $P_{e\parallel} \propto r^3$, holds then $\Gamma_e=0$ and the plasma becomes isotropic ($P_{e\parallel}=P_{e\perp}$) otherwise for other Landau levels $P_{e\parallel}\neq P_{e\perp}$, in general.

Now let us consider a collisionless quasineutral uniformly magnetized Fermi-Dirac degenerate plasma consisting of mobile electrons and ions. The arbitrary strength uniform magnetic field, ${\bf{B}}$, is assumed to be directed along the $x$ axis. The full quantum hydrodynamics equations describing the dynamics of electrons and ion, disregarding the Bohm-force for simplicity, take the form
\begin{equation}\label{dimensional}
\begin{array}{l}
\frac{{\partial {n_{i,e}}}}{{\partial t}} + \nabla \cdot({n_{i,e}}{{\bf{v_{i,e}}}}) = 0,{{\bf{v_{(i,e)}}}} = {\bf{\hat i}}{v_{x(i,e)}} + {\bf{\hat j}}{v_{y(i,e)}} + {\bf{\hat k}}{v_{z(i,e)}}, \\
\frac{{\partial {{\bf{v_i}}}}}{{\partial t}} + ({{\bf{v_i}}}\cdot\nabla ){{\bf{v_i}}} + \frac{e}{{{m_i}}}\nabla \phi  + \frac{1}{{{m_i}{n_i}}}\nabla  \cdot {{{\bf{\tilde P}}}_i}{\rm{(}}{{\rm{n}}_{\rm{i}}}{\rm{)}} - {\omega _{ci}}({{\bf{v_i}}} \times {\bf{\hat i}}) = 0, \\
\frac{{{m_e}}}{{{m_i}}}\left( {\frac{{\partial {{\bf{v}}_e}}}{{\partial t}} + ({{\bf{v}}_e}\cdot\nabla ){{\bf{v}}_e} + {\omega _{ce}}({{\bf{v}}_e} \times {\bf{\hat i}})} \right) + \frac{1}{{{m_i}{n_e}}}+ \nabla  \cdot {{{\bf{\tilde P}}}_e}{\rm{(}}{{\rm{n}}_{\rm{e}}}{\rm{)}} = \frac{e}{{{m_i}}}\nabla \phi , \\
\Delta \phi  = 4\pi e({n_e} - {n_i}) \approx 0, \\
\end{array}
\end{equation}
where, $v_{(i,e)}$, $m_{(i,e)}$ and $\omega_{c{(i,e)}}=eB/m_{(i,e)}$ are the ion/electron velocity, mass and cyclotron frequencies, respectively. Also, ${{{\bf{\tilde P}}}_e}{\rm{(}}{{\rm{n}}_{\rm{e}}}{\rm{)}}$ and ${{{\bf{\tilde P}}}_i}{\rm{(}}{{\rm{n}}_{\rm{i}}}{\rm{)}}$ are the electron and ion diagonal pressure tensor, given as \cite{cheong, mahmood}
\begin{equation}\label{pt}
{{\bf{\tilde P}}_e}({{\rm{n}}_e}){\rm{ = }}\left[ {\begin{array}{*{20}{c}}
   {{P_{e\parallel }}} & 0 & 0  \\
   0 & {{P_{e \bot }}} & 0  \\
   0 & 0 & {{P_{e \bot }}}  \\
\end{array}} \right],\hspace{3mm}{{\bf{\tilde P}}_i}({{\rm{n}}_i}){\rm{ = }}\left[ {\begin{array}{*{20}{c}}
   {{P_{i\parallel }}} & 0 & 0  \\
   0 & {{P_{i \bot }}} & 0  \\
   0 & 0 & {{P_{i \bot }}}  \\
\end{array}} \right]
\end{equation}
It is noted that the mathematical treatment employed here have quite resemblance to the CGL theory of double-adiabatic ion-pressure \cite{bau, parks} where ${\bf{\tilde P}} = {P_ \bot }{\bf{\tilde I}} + ({P_\parallel } - {P_ \bot }){\bf{\hat b\hat b}}$, with $\bf{\hat b}$ being the unit vector along the nonuniform ambient magnetic-field. In the case where the magnetic-field is anchored to the plasma density such as for frozen-in-field or Landau-orbital-ferromagnetism (LOFER) \cite{can4} the nondiagonal elements of stress-tensor are nonzero and the pressure-tensor can be written in a more general form \cite{krall}
\begin{equation}\label{pk}
{\bf{\tilde P}}{\rm{ = }}\left[ {\begin{array}{*{20}{c}}
   {{P_ \bot } + ({P_\parallel } - {P_ \bot }){b_x}{b_x}} & {({P_\parallel } - {P_ \bot }){b_x}{b_y}} & {({P_\parallel } - {P_ \bot }){b_x}{b_z}}  \\
   {({P_\parallel } - {P_ \bot }){b_y}{b_x}} & {{P_ \bot } + ({P_\parallel } - {P_ \bot }){b_y}{b_y}} & {({P_\parallel } - {P_ \bot }){b_y}{b_z}}  \\
   {({P_\parallel } - {P_ \bot }){b_z}{b_x}} & {({P_\parallel } - {P_ \bot }){b_z}{b_y}} & {{P_ \bot } + ({P_\parallel } - {P_ \bot }){b_z}{b_z}}  \\
\end{array}} \right]
\end{equation}
In order to obtain a dimensionless set of equations, we use general scaling defined below
\begin{equation}\label{nm}
\nabla \to \frac{1}{\lambda_i}\bar \nabla,\hspace{2mm}t \to \frac{{\bar t}}{{{\omega _{pi}}}},\hspace{2mm}{n_{(i,e)}} \to {n_0}{\bar n_{(i,e)}}, \hspace{2mm}{\bf{v_{(i,e)}}} \to {c_i}{\bf{\bar v_{(i,e)}}},\hspace{2mm}\phi  \to \frac{\epsilon}{e}\bar \phi,\hspace{2mm}P_{(i,e)} \to \epsilon \bar P_{(i,e)},
\end{equation}
where, $\omega_{pi}=\sqrt{4\pi e^2 n_0/m_i}$ is the plasma ion frequency, ${\lambda_i} = c_i/\omega_{pi}$ the ion gyroradius, and $c_i=\sqrt{\epsilon/m_i}$ the ion sound-speed, values of which  will be defined later along with the parameter $\epsilon$ based on the quantum statistical distribution of electrons. Furthermore, assuming that $m_i/m_e \gg 1$, the dimensionless set of equations, disregarding the bar notation, take the following simplified form
\begin{equation}\label{normal2}
\begin{array}{l}
\frac{{\partial n}}{{\partial t}} + \nabla \cdot(n{\bf{v_i}}) = 0, \\
\frac{{\partial {\bf{v_i}}}}{{\partial t}} + ({\bf{v_i}}\cdot\nabla ){\bf{v_i}} + \frac{{1}}{n}\nabla  \cdot {{{\bf{\tilde P}}}}{\rm{(}}{{\rm{n}}_{\rm{}}}{\rm{)}} - \bar\omega({\bf{v_i}} \times {\bf{\hat i}}) = 0,\\ {\bf{\tilde P}}({\rm{n}}) = {{\bf{\tilde P}}_{\bf{i}}}({{\rm{n}}_i}) + {{\bf{\tilde P}}_{\bf{e}}}({{\rm{n}}_{\rm{e}}}). \\
\end{array}
\end{equation}
where, $\bar\omega=\omega_{ci}/\omega_{pi}$ is the normalized magnetic field strength. Taking $\bf{v_i}=\bf{v}$ in this scheme, we further simplify the model to a two-dimensional perturbation in $x$-$y$ plane and ignoring the ion pressure regarding that of the electron degeneracy. Note that in the ground-state Landau-level where the transverse electron degeneracy pressure vanishes the presence of ion-pressure becomes essential and we will consider the ion pressure when we get back to this special case in later sections. The simplified set of equations regarding the dynamics of ion-waves are as follows
\begin{equation}\label{scalar}
\begin{array}{l}
{\partial _t}n + {\partial _x}(n{v_x}) + {\partial _y}(n{v_y}) = 0, \\
{\partial _t}{v_x} + \left( {{v_x}{\partial _x} + {v_y}{\partial _y}} \right){v_x} + \frac{{{\partial _x}P_{\parallel}(n)}}{n} = 0, \\
{\partial _t}{v_y} + \left( {{v_x}{\partial _x} + {v_y}{\partial _y}} \right){v_y} + \frac{{{\partial _y}P_{\perp}(n)}}{n} - \bar\omega{v_z} = 0, \\
{\partial _t}{v_z} + \left( {{v_x}{\partial _x} + {v_y}{\partial _y}} \right){v_z} + \bar\omega{v_y} = 0. \\
\end{array}
\end{equation}
The Eqs. (\ref{scalar}) may be solved together to obtain the nonlinear evolution of magneto-ion-acoustic waves in an anisotropic quantum plasma a task which will be followed in next section.

\section{General Anisotropic Soliton Solution}\label{Sagdeev}

In order to obtain a stationary soliton solution to equation-set we use the transformation $\xi=k_\parallel x+k_\perp y-M t$, where, $M=V/c_i$ is the Mach-number of solitons. Furthermore, by a trivial integration with the boundary conditions ($\mathop {\lim }\limits_{v_x,v_y  \to 0} n = 1$), the Eqs. (\ref{scalar}) reduce to the form
\begin{equation}\label{red}
\begin{array}{l}
\left( { - M + {k_\parallel}{v_x} + {k_\perp}{v_y}} \right) = - M/n, \\
\left( { - M + {k_\parallel}{v_x} + {k_\perp}{v_y}} \right){d_\xi }{v_x} + \frac{{{k_\parallel}}}{n}{d_\xi }P_{\parallel}(n) = 0, \\
\left( { - M + {k_\parallel}{v_x} + {k_\perp}{v_y}} \right){d_\xi }{v_y} + \frac{{{k_\perp}}}{n}{d_\xi }P_{\perp}(n) - \bar\omega{v_z} = 0, \\
\left( { - M + {k_\parallel}{v_x} + {k_\perp}{v_y}} \right){d_\xi }{v_z} + \bar\omega{v_y} = 0. \\
\end{array}
\end{equation}
By standard mathematical procedure the reduced equation-set Eqs. (\ref{red}) further simplify to the following single differential equation
\begin{equation}\label{sol}
\begin{array}{l}
\frac{d}{{d\xi }}\left\{ {\frac{1}{n}\left[ {\frac{{{d^2}}}{{d{\xi ^2}}}\left( {\frac{{{M^2}}}{{2{n^2}{{\bar \omega }^2}}} + {\Psi _1}(n)} \right) + 1} \right]} \right\} + \frac{n}{{{M^2}}}\frac{{d{\Psi _2}(n)}}{{d\xi }} = 0, \\
{\Psi _1}(n) = k_\parallel ^2\int_1^n {\frac{{{d_n}{P_\parallel }(n)}}{n}dn + } k_ \bot ^2\int_1^n {\frac{{{d_n}{P_ \bot }(n)}}{n}dn},\hspace{3mm}{\Psi _2}(n) = k_\parallel ^2\int_1^n {\frac{{{d_n}{P_\parallel }(n)}}{n}dn},  \\
\end{array}
\end{equation}
where, $\Psi (n)$'s are the generalized effective potentials. Using some algebraic manipulation assuming the aforementioned boundary conditions lead to the well-known energy integral of the form
\begin{equation}\label{energy}
\frac{1}{2}{\left( {\frac{{dn}}{{d\xi }}} \right)^2} + U(n) = 0, \\
\end{equation}
where the generalized pseudo-potential for an anisotropic quantum plasma is given as
\begin{equation}\label{pseudo}
\begin{array}{l}
\begin{array}{l}
U(n) = {\left[ {{d_n}{\Psi _1}(n) - \frac{{{M^2}}}{{{n^3}{{\bar \omega }^2}}}} \right]^{ - 2}} \times  \\
\int_1^n {\left\{ {\left[ {{d_{n'}}{\Psi _1}(n') - \frac{{{M^2}}}{{{{n'}^3}{{\bar \omega }^2}}}} \right]\left[ {1 - n' + \frac{{n'}}{{{M^2}}}\int_1^{n'} {n''{d_{n''}}{\Psi _2}(n'')dn''} } \right]} \right\}dn'}  \\
\end{array}
\end{array}
\end{equation}
It is easily confirmed that, the Sagdeev pseudopotential given in Eq. (\ref{pseudo}) satisfies the first two conditions below
\begin{equation}\label{conditions}
{\left. {U(n)} \right|_{n = 1}} = {\left. {\frac{{dU(n)}}{{dn}}} \right|_{n = 1}} = 0,\hspace{3mm}{\left. {\frac{{{d^2}U(n)}}{{d{n^2}}}} \right|_{n = 1}} < 0.
\end{equation}
On the other hand, the third condition in Eq. (\ref{conditions}) leads to
\begin{equation}\label{con2}
{\left. {\frac{{{d^2}U(n)}}{{d{n^2}}}} \right|_{n = 1}} = \frac{{\Delta _1^2{{\bar \omega }^2}}}{{\Delta _2^2{M^2}}}\left[ {\frac{{{M^2} - \Delta _2^2}}{{{M^2} - \Delta _1^2{{\bar \omega }^2}}}} \right] < 0,
\end{equation}
or, for the stable Mach-range we obtain
\begin{equation}\label{mr}
\left\{ {\begin{array}{*{20}{c}}
{{\Delta _2} < M < \bar \omega {\Delta _1}} & {\frac{{{\Delta _2}}}{{{\Delta _1}}} < \bar \omega }  \\
{{\Delta _2} > M > \bar \omega {\Delta _1}} & {\frac{{{\Delta _2}}}{{{\Delta _1}}} > \bar \omega }  \\
\end{array}} \right\},
\end{equation}
where
\begin{equation}\label{con3}
{\Delta _1} = \sqrt {{{\left. {{d_n}{\Psi _1}(n)} \right|}_{n = 1}}},\hspace{3mm}{\Delta _2} = \sqrt {{{\left. {{d_n}{\Psi _2}(n)} \right|}_{n = 1}}}.
\end{equation}
In the following sections we apply the criteria obtained to different situations in magnetized degenerate plasmas and show that many different plasma parameters can affect the stability of the localized density perturbations in quantum plasmas.

\section{Quantum Double-Degenerate Plasmas}

\subsection{Fermi Plasmas and Dimensionality Effect}\label{discussion1}

For a normally degenerate Fermi electron-gas the normalized parallel pressure components are of the form $P_{e\parallel}=2n^{(d+2)/d}/(2+d)$ (with $d=2,3$ being the system dimensionality) and the normalized ion-pressure can be written as $P_{i\parallel}=\sigma_{\parallel} n^3$ \cite{mahmood} with $\sigma_{\parallel}=T_{i\parallel}/T_{Fe} \ll 1$, where, $T_{Fe}$ is the electron Fermi-temperature and $d$ is the system dimensionality. Note that the normalizing parameters are taken as $\epsilon=2k_B T_{Fe}$ and $c_i=\sqrt{2k_B T_{Fe}/m_i}$. Therefore, including the ion pressure also we obtain, $\Delta_2=\cos\theta\sqrt{2/d+3\sigma_{\parallel}}$, where, $\theta$ is the angle of propagation with respect to the external magnetic-field. On the other hand, the magnetic susceptibility of a weakly magnetized ($B\mu_B\ll k_B T_{Fe}$) Fermi gas is $\chi=\mu_B^{2} D(E_{Fe})(1-\mu^2/3)$ \cite{sho}, where, $\mu_B=e\hbar/2m_e c$ is the Bohr-magneton, $D(E_{Fe})=nd/2E_{Fe}$ the electron density-of-states, and $\mu=m_e/m_e^{*}$ is the effective electron mass ratio. Therefore, using the relation $P_{e\perp}=P_{e\parallel}-B\Gamma_e$, with $\eta=B \mu_B/2E_{Fe}$ being the Zeeman-energy parameter, we obtain in normalized form $P_{e\perp}=P_{e\parallel}-\eta^2 nd (1-\mu^2/3)$ and from CGL theory $P_{i\perp}=\sigma_{\perp}n$ ($\sigma_{\perp}=T_{i\perp}/T_{Fe} \ll 1$) which leads to $\Delta_1=\sqrt{2/d+3\sigma_{\parallel}\cos^2\theta+\sigma_{\perp}\sin^2\theta-d\eta^2(1-\mu^2/3)\sin^2\theta}$. It can be observed that the forth term in $\Delta_1$ can change the sign depending on the critical value $\mu_{cr}=\sqrt{3}$. It has been shown to lead to distinct nonlinear features in paramagnetic quantum plasmas \cite{akbari10}, since the value of $\mu$ can be as large as $10^3$ for some Bismuth compounds \cite{zhang} or as low as $10^{-3}$ for heavy fermion semiconductors. Neglecting the ion pressure effects one observes that the magnetization induced pressure anisotropy can be detected for $d=3, \mu=1$ with the condition $B\simeq 2.6\times10^{-7} n_e^{2/3}$ and for $d=2,\mu=1$ with the condition $B\simeq 3.6\times10^{-6} n_e^{2}$. It is obvious that when the value of fractional effective mass becomes negligibly small, the laboratory scale magnetic fields may produce a measurable magnetization-induced pressure anisotropy effects, so that, $\Delta_1$ differs significantly from $\Delta_2$.

\subsection{Finite-Temperature Thomas-Fermi Plasma}\label{discussion0}

\textbf{For the Thomas-Fermi degenerate plasma with trapped electrons the normalized electron density is of the form $n_e=(1+\phi)^{3/2}+T_e^2(1+\phi)^{-1/2}$ \cite{shah} and the parallel component of the classical ion pressure is again $P_{i\parallel}=\sigma_{\parallel} n^3$ with $\sigma_{\parallel}=T_{i\parallel}/T_{Fe} \ll 1$ and the normalization parameters, $\epsilon$ and $c_i$ as in the previous case. We obtain for this case $\Delta_2=\cos\theta\sqrt{2/(3-T_e^2)+3\sigma_{\parallel}}$. The quantum paramagnetic susceptibility, on the other hand, in the weak-field limit ($\eta\ll 1$) is given by the Curie-Law, $\chi=C/T_e$ with $C=n_e p^2 \mu_B^2/3$ being the Curie-constant and $p$ being the number of Bohr magnetons \cite{kittel}. Therefore one obtains $\Delta_1=\sqrt{2/(3-T_e^2)+3\sigma_{\parallel}\cos^2\theta+\sigma_{\perp}\sin^2\theta- 2\eta^2 T_{Fe}\sin^2\theta/3T_e}$ ($T_e\ne 0$). In the ferromagnetic-limit one must use the Curie-Weiss law ($\chi=C/(T_e-T_c)$ with $T_c$ being the critical temperature) and consequently at very low temperatures when the electron-temperature, $T_e$, approaches the critical temperature, $T_c$, the negative term in the expression for $\Delta_1$ may dominate over the other terms leading to detectable nonlinear double degeneracy instability for some propagation angles, $\theta$.}

\subsection{Relativistically Degenerate Plasmas}\label{discussion2}

For a relativistically degenerate Fermi-Dirac electron-gas the normalized pressures can be written as $P_e(r)=\frac{1}{8r_0^{3}}\left\{ {{r}\left( {2{r^{2}} - 3} \right)\sqrt {1 + {r^{2}}}  + 3\text{sinh}^{-1}{r}} \right\}$ \cite{chandra1}, where, $r=r_0 n^{1/3}$ is the relativity parameter and $r_0=(n_{0}/n_c)^{1/3}$ (${n_c} = \frac{{8\pi m_e^{3}{c^3}}}{{3{h^3}}}\simeq 5.9 \times 10^{29} cm^{-3}$) is the relativistic degeneracy parameter. Furthermore, in this case the normalized ion-pressure which is very small can be written as $P_{i}=\sigma n^s$ (where, the parameter $s$ is defined through the CGL theory) with $\sigma=k_B T_i/m_e c^2 \simeq 0$ with the normalizing parameters taken as $\epsilon=m_e c^2$ and $c_i=c\sqrt{m_e/m_i}$. Thus, we have $\Delta_2=r_0({1 + r_0^2})^{-1/4}\cos\theta /\sqrt{3}$. On the other hand, the weak-filed magnetic susceptibility for a relativistically degenerate Fermi-Dirac plasma in normalized form is given as, $\chi_r\simeq 3n\mu_B^{2}\sqrt{1+r^2} / m_e^{2} c^4 r^{2}$. This gives rise to $\Delta_1=\sqrt{r_0^{2}({1 + r_0^2})^{-1/2}/3-3\eta^{2}\sqrt{1+r_0^{2}} \sin^2\theta/r_0^{2}}$, again with $\eta=B\mu_B/m_e c^2$ being the Zeeman-energy parameter. It is clearly remarked that the competing terms in $\Delta_1$ can interplay in low-density and high magnetic-field ($3\eta\simeq r_0^2\ll 1$ or $B\simeq4.2\times10^{-7} n_e^{2/3}$) degenerated plasma leading to large magnetization pressure anisotropy effects. Note that the condition on magnetic field and density relation in this case is quite close to the one given above.

\subsection{Coupled Relativistically Degenerate Plasmas}\label{discussion2}

In a relativistically degenerate plasma due to decrease in inter-fermion distances the coupling parameter may exceed unity. In such situations the Thomas-Fermi non-uniformity in electron distribution, Coulomb effect, electron exchange and ion-correlations introduce minor corrections to the Chandrasekhar electron degeneracy pressure of which the contribution due to the Coulomb interactions is the largest \cite{salpeter}. The negative Coulomb plus the Thomas-Fermi screening pressure, reads as
\begin{equation}\label{c}
{P_{C + TF}} =  - \frac{{8{\pi ^3}{m_e^4}{c^5}}}{{{h^3}}}\left[ {\frac{{\alpha {Z^{2/3}}}}{{10{\pi ^2}}}{{\left( {\frac{4}{{9\pi }}} \right)}^{1/3}}{r^4} + \frac{{162}}{{175}}\frac{{{{(\alpha {Z^{2/3}})}^2}}}{{9{\pi ^2}}}{{\left( {\frac{4}{{9\pi }}} \right)}^{2/3}}\frac{{{r^5}}}{{\sqrt {1 + {r^2}} }}} \right].
\end{equation}
where, the parameters, $\alpha=e^2/\hbar c\simeq 1/137$ and $Z$ are the fine-structure constant and the atomic number, respectively. Therefore, in the same normalization scheme as above and ignoring the $\alpha^2$ terms, one may write
\begin{equation}\label{eff2}
{\Psi _C}(n) = \int {\frac{1}{{n}}} \frac{{d{P_C}(n)}}{{dn}}dn=\beta r_0 n^{1/3},
\end{equation}
where, $\beta=\alpha ({2^{5/3}}/5){(3{Z^2}/\pi )^{1/3}}$. Therefore, for the strongly-coupled magnetized Fermi-Dirac plasma considered here, we obtain $\Delta_2=\cos\theta \sqrt{r_0^{2}({1 + r_0^2})^{-1/2}-\beta r_0}/\sqrt{3}$ and ${\Delta _1} = \sqrt {({r_0}{{(1 + r_0^2)}^{ - 1/2}} - \beta ){r_0}/3 - 3{\eta ^2}\sqrt {1 + r_0^{2}} {{\sin }^2}\theta/r_0^{2} }$, again with $r_0=(n_{0}/n_c)^{1/3}$ and $\eta=B\mu_B/m_e c^2$ being the relativistic degeneracy and the Zeeman-energy parameters. More recently, it has been shown that in a magnetized quantum plasma the Coulomb force can be comparable to that of degeneracy and the Coulomb instability may be set-up in laboratory conditions for $r_0\simeq 0.01$ ($n_0\simeq 10^{23}/cm^{3}$) and $\eta\simeq 10^{-8}$ ($B\simeq 100T$) \cite{akbarinew}. In such case the first term in ${\Delta _1}$ can be compared to the second term due the plasma magnetization. Therefore, there may be situations detectable in laboratory quantum plasmas where the magnetization pressure-anisotropy can play a role.

\subsection{Landau-Quantized Degenerate Plasmas}\label{discussion2}

It has been shown \cite{can3} that the energy spectrum of a magnetized degenerate plasma is quantized due to the electron spin-orbit magnetization. The thermodynamic quantities are also quantized due to quantization of electronic density of states. In a zero-temperature Fermi-gas the pressure is highly anisotropic and the gas may become one-dimensional along the external field in the quantum-limit ($l=0$) \cite{can4}. Plasma density, parallel pressure to the field can be analytically expressed in terms of Hurwitz zeta-functions as below \cite{claud}
\begin{equation}\label{h}
\begin{array}{l}
n_e(r',\gamma ) = {n_c}{(2\gamma )^{3/2}}{H_{ - 1/2}}\left( {\frac{{{r'^2}}}{{2\gamma }}} \right), \\
{P_{e\parallel}}(r',\gamma ) = \frac{{{n_c}{m_e}{c^2}}}{2}{(2\gamma )^{5/2}}\int_{0}^{\frac{{{r'^2}}}{{2\gamma }}} {\frac{{{H_{ - 1/2}}(q)}}{{\sqrt {1 + 2\gamma q} }}} dq, \\
{H_\nu}(q) = h(\nu,\{ q\} ) - h(\nu, q + 1 ) - \frac{1}{2}{q^{ - \nu}}, \\
h(\nu,q) = \sum\limits_{n = 0}^\infty  {{{(n + q)}^{ - \nu}}} . \\
\end{array}
\end{equation}
where $h(\nu,\{q\})$ is the Hurwitz zeta-function of order $\nu$ with the fractional part of $q$ as argument and $n_c=m_e^3 c^3/2\pi^2 \hbar^3$, $\gamma=B/B_c$ with $B_c=m_e^2c^3/e\hbar\simeq4.414\times 10^{13}G$ are the density normalization and fractional critical-field parameters and $r'=P_{Fe}/m_e c$ is the relativity parameter. Note that the definition of $n_c$ in this case is different from that defined earlier. In the limit $r'^2/2\gamma\gg 1$ the plasma pressure reduces to that of the Chandrasekhar and the plasma becomes isotropic, while, for $r'^2/2\gamma\ll 1$ the plasma is in its ground-state (spin-orbit) quantization level where $P_{e\perp}=P_{e\parallel}-B\Gamma_{e}=0$, where, $\Gamma_{e}$ is the electron spin-orbit magnetization. Therefore, one may use the thermodynamic relations $n_e(r',\gamma ) = c_{s}^{-2}\partial {P_{e\parallel}(r',\gamma )}/\partial {\epsilon _{Fe}}$ ($\epsilon _{Fe}=\sqrt{1+r'^2}$) and $\Gamma_{e}(r',\gamma ) = B_c^{-1}{{\partial {P_{e\parallel}}(r',\gamma )}}/{{\partial \gamma}}$ for a homogenous magnetized plasma to numerically evaluate the Much-number range available for solitary wave propagation.

From Eq. (\ref{con3}), it is evident that for a Landau-quantized plasma, we have $\Delta_1=\Delta_2/k_{\parallel}$ in the classical limit ($l=\infty$), i.e. for isotropic plasmas ($P_{e\parallel}=P_{e\perp}$). In general, the following Mach-number limits is obtained for solitary wave propagation in the classical-limit
\begin{equation}\label{con4}
\left\{ {\begin{array}{*{20}{c}}
{{k_\parallel } < \frac{M}{{\sqrt {{{\left. {{d_n}{P_{e\parallel}}(n)} \right|}_{n = 1}}} }} < \bar \omega ;} & {\bar \omega  > {k_\parallel }}  \\
{\bar \omega  < \frac{M}{{\sqrt {{{\left. {{d_n}{P_{e\parallel}}(n)} \right|}_{n = 1}}} }} < {k_\parallel };} & {{k_\parallel } > \bar \omega }  \\
\end{array}} \right\}.
\end{equation}

Moreover, in the quantum-limit ($l=0$) where $P_{e\perp}=0$, we are led to $\Delta_1=\Delta_2$ and arrive at the following Mach-limits for the stability of solitary excitations
\begin{equation}\label{con5}
\left\{ {\begin{array}{*{20}{c}}
{1 < \frac{M}{{{k_\parallel }\sqrt {{{\left. {{d_n}{P_{e\parallel}}(n)} \right|}_{n = 1}}} }} < \bar \omega ;} & {\bar \omega  > 1}  \\
{\bar \omega  < \frac{M}{{{k_\parallel }\sqrt {{{\left. {{d_n}{P_{e\parallel}}(n)} \right|}_{n = 1}}} }} < 1;} & {1 > \bar \omega }  \\
\end{array}} \right\}.
\end{equation}
It is concluded that, the two limits coincide for parallel propagation ($k_{\parallel}=1$), while, the perpendicular propagation ($k_{\parallel}=0$) for the quantum limit unlike the classical limit is prohibited.

\textbf{The formulation presented in this research can also be further extended to include the pair-ion, electron-positron-ion, dusty, two-electron-temperature, finite-temperature quantum plasmas as well as multi-species classical plasmas. As it is evident, the direct experimental evidence for the theory presented here requires a sustainable high magnetic field which may be accessible via rapidly growing field of strong laser-matter interactions in near future \cite{mend2}.}

\section{Concluding Remarks}\label{conclusion}

We used a double-degeneracy quantum hydrodynamics formalism analogous to the well-known double-adiabatic theory to evaluate the possibility of a nonlinear localized-density excitations in anisotropic magnetized quantum plasmas. The application to different situation was presented and found that the parameters such as dimensionality, ion-temperature, relativistic-degeneracy, Zeeman-energy, and plasma composition can alter the stability of oblique solitary wave propagation. The analysis presented here can be applied to a variety
of environments particularly to astrophysical dense stellar structure such as white-dwarfs, neutron-star crusts and pulsar magnetospheres.


\begin{thebibliography}{}

\bibitem{davidson} R. C. Davidson, "Methods in Nonlinear Plasma Theory", Academic Press, New York, (1972).
\bibitem{vedenov} A. A. Vedenov, E. P. Velikhov, and R. Z. Sagdeev, Nuclear Fusion, \textbf{1}, 82(1961)[in Russian].
\bibitem{sagdeev} R. Z. Sagdeev, 1966 "Reviews of Plasma Physics", Vol. 4 ed. M. A. Leontovich (NewYork: Consultants Bureau).
\bibitem{ikezi} H. Ikezi, R. Taylor and D. Baker, Phys. Rev. Lett. 25 11(1970).
\bibitem{lee} L. C. Lee and J. R. Kan, Phys. Fluids \textbf{24}, 430 (1981).
\bibitem{witt} E. Witt and W. Lotko, Phys. Fluids \textbf{26}, 2176 (1983).
\bibitem{yashvir} Yashvir, T. N. Bhatnagar and S. R. Sharma , Plasma Phys. Control. Fusion \textbf{26}, 1303(1984).
\bibitem{yadav} L. L. Yadav and S. R. Sharma, Physics Letters A, \textbf{150}, 397(1990).
\bibitem{cheong} Cheong Rim Choi, Chang-Mo Ryu, D.-Y.Lee, Nam C. Lee, Y.-H. Kim, Phys. Lett. A, \textbf{364} 297(2007).
\bibitem{mahmood} S Mahmood, S Hussain, W Masood, and H Saleem, Phys. Scr., \textbf{79} 045501(2009).
\bibitem{chew} G. F. Chew, M. L. Goldberger and F. E. Low, Proc. Roy. Soc (London), \textbf{A236} 112(1956).
\bibitem{can0} V. Canuto and Chih Kang Chuo, Phys. Fluids, \textbf{16}, 1273(1973).
\bibitem{akbari00} M. Akbari-Moghanjoughi, Phys. Plasmas, \textbf{18} 084701(2011).
\bibitem{stan} P. K. Shukla and L. Stenflo, J. Plasma Phys. \textbf{74} 575(2008).
\bibitem{bohm} D. Bohm and D. Pines, Phys. Rev. \textbf{92} 609(1953).
\bibitem{pines} D. Pines, Phys. Rev. \textbf{92} 609(1953).
\bibitem{levine} P. Levine and O. V. Roos, Phys. Rev, \textbf{125} 207(1962).
\bibitem{landau} L. D. Landau and E. M. Lifshitz, "Statistical Physics", Part I, Pergamon, Oxford, (1978).
\bibitem{sho} D. Shoenburg, Phil. Trans. R. Soc. Lond., \textbf{A255} 85(1962).
\bibitem{haas1} G. Manfredi and F. Haas, Phys. Rev. B \textbf{64}, 075316 (2001).
\bibitem{haas2} F. Haas, L. G. Garcia, J. Goedert, and G. Manfredi, Phys. Plasmas \textbf{10}, 3858(2003).
\bibitem{Markowich} P. A. Markowich, C. A. Ringhofer, and C. Schmeiser, Semiconductor Equations (Springer-Verlag, New York, 1990).
\bibitem{Marklund1} M. Marklund and P. K. Shukla, Rev. Mod. Phys. \textbf{78}, 591 (2006).
\bibitem{Brodin1} G. Brodin, M. Marklund, and G. Manfredi, Phys. Rev. Lett. \textbf{100}, 175001 (2008).
\bibitem{Marklund2} M. Marklund, G. Brodin, L. Stenflo, and C. S. Liu, Europhys. Lett. \textbf{84}, 17006 (2008).
\bibitem{Brodin2} G. Brodin and M. Marklund, New J. Phys. \textbf{9}, 277(2007).
\bibitem{manfredi} G. Manfredi, Fields Inst. Commun. \textbf{46} 263(2005).
\bibitem{shukla} Shukla P K, Eliasson B, "Nonlinear aspects of quantum plasma physics" Phys. Usp. \textbf{51} 53(2010).
\bibitem{gardner} C. Gardner, SIAM, J. Appl. Math. \textbf{54} 409(1994).
\bibitem{haug} H. Haug and S. W. Koch, "Quantum theory of the optical and electronic properties of semiconductors", World Scientific, (London)2004,
\bibitem{chandra1} S. Chandrasekhar, "An Introduction to the Study of Stellar Structure", Chicago, Ill. , (The
University of Chicago press), (1939).
\bibitem{chandra2} S. Chandrasekhar, Mon. Not. R. Astron. Soc., \textbf{113} 667(1953).
\bibitem{chandra3} S. Chandrasekhar, Science, \textbf{226} 4674(1984).
\bibitem{kothary} D. S. Kothari and B. N. Singh,  Proceedings of the Royal Society of London. Series A, Mathematical and Physical
Sciences, Vol. 180, No. 983 (Jul. 3, 1942), pp. 414-423
\bibitem{can1} V. Canuto, H. Y. Chiu, and L. Fassio-Canuto, Astrophys. Space Sci. \textbf{3} 258(1969).
\bibitem{can2} Vittorio Canuto and Hong Yee Chiu, Phys. Rev., \textbf{173} 1229(1968).
\bibitem{can3} Hong Yee Chiu and Vittorio Canuto, Phys. Rev. Lett., \textbf{21} 110(1968).
\bibitem{can4} Hyung Joon Lee, Vittorio Canuto, Hong Yee Chiu and Claudio Chiuderi, Phys. Rev. Lett., \textbf{23} 390(1969).
\bibitem{con1} R. F. O'Connell and K. M. Roussel, Astron. Astrophys., \textbf{18} 198(1972).
\bibitem{con2} R. F. O'Connell and K. M. Roussel, Nat. Phys. Sci., \textbf{231} 32(1971).
\bibitem{con3} R. F. O'Connell and K. M. Roussel, LETTERE AL NUOVO CIMENTO, \textbf{2} 815(1971).
\bibitem{burk1} Johannes Schmid-Burgk, Astron. Astrophys., \textbf{26} 335(1973).
\bibitem{burk2} H. Pohl and J. Schmid-Burgk,, Nat. Phys. Sci., \textbf{238} 56(1972).
\bibitem{Marklund3} M. Marklund and G. Brodin, Phys. Rev. Lett., \textbf{98} 025001(2007).
\bibitem{Marklund4} M. Marklund and B. Eliasson and P. K. Shukla, Phys. Rev. E., \textbf{76} 067401(2007).
\bibitem{misra1} A. P. Misra and P. K. Shukla, Phys. Plasmas, \textbf{15} 052105(2010).
\bibitem{martin} Martin Stefan, Gert Brodin and Mattias Marklund, New J. Phys., \textbf{12} 013006(2010).
\bibitem{mushtaq} A. Mushtaq, and S. V. Vladimirov, Phys. Plasmas, \textbf{17} 102310(2010).
\bibitem{brodin3} G. Brodin and M. Marklund, Phys. Plasmas, \textbf{14} 112107(2007).
\bibitem{zaman} J. Zamanian, G. Brodin and M. Marklund, New J. Phys., \textbf{11} 072017(2009).
\bibitem{vitaly} Vitaly Bychkov, Mikhail Modestov, and Mattias Marklund, Phys. Plasmas, \textbf{17} 112107(2010).
\bibitem{misra2} A. P. Misra, G. Brodin, M. Marklund, and P. K. Shukla, Phys. Plasmas, \textbf{17} 122306(2010).
\bibitem{akbari0} M. Akbari-Moghanjoughi, Phys. Plasmas, \textbf{18} 112708(2011).
\bibitem{crut} Richard M. Crutcher, The Astrophys. J., \textbf{520} 706(1991).
\bibitem{kemp} J. C. Kemp, J. B. Swedlund, J. D. Landstreet and J. R. P. Angel, The Astrophys. J. \textbf{L77} 161(1970).
\bibitem{put} A. Putney, The Astrophys. J. \textbf{L67} 451(1995).
\bibitem{jor} S. Jordan, Astr. \& Astrophys. \textbf{265} 570(1992).
\bibitem{dong} Dong Lai, Rev. Mod. Phys., \textbf{73} 629(2001).
\bibitem{harding} Alice K. Harding and Dong Lai, Rep. Prog. Phys., \textbf{69} 2631(2006) Doi: 10.1088/0034-4885/69/9/R03
\bibitem{chai} M. Chaichian, S. S. Masood, C. Montonen, A. P\'{e}rez Mart\'{\i}nez, and H. P\'{e}rez Rojas, Phys. Rev. Lett., \textbf{84} 5261(2000).
\bibitem{akbari7} M. Akbari-Moghanjoughi, Phys. Plasmas, \textbf{18} 112708(2011).
\bibitem{suh} In-Saeng Suh and G. J. Mathews, The Astrophys. J., \textbf{530} 949(2000).
\bibitem{salpeter} E. E. Salpeter, APJ., \textbf{134} 669(1961). "http://adsabs.harvard.edu/full/1961ApJ...134..669S"
\bibitem{mend1} Shalom Eliezer, Peter Norreys, Jos\'{e} T. Mendon\'{c}a and Kate Lancaster, Phys. Plasmas, \textbf{12} 052115(2005).
\bibitem{mend2} S. Eliezer, P. A. Norreys, J. T. Mendon\'{c}a, K. L. Lancaster, Central Laser Facility Ann. Rep. (High Power Laser Programme – Theory and Computation) (2004/2005) pp.99.
\bibitem{claud} Claudio O. Dib and Olivier Espinosa, Nucl. Phys. B \textbf{612} 492(2001).
\bibitem{bau} Baumjohann W and Treumann R A, "Basic Space Plasma Physics" (London: Imperial College Press), 1997.
\bibitem{parks} Parks G K, "Physics of Space Plasmas" (USA: Perseus), 1991.
\bibitem{krall} Nicholas A Krall and Alvin W Trivelpeice, "Principles of Plasma Physics" (USA: McGraw-Hill Book Company), 1932.
\bibitem{akbari10} M. Akbari-Moghanjoughi, IEEE Trans. Plasma Sci., \textbf{39} 3180(2011).
\bibitem{zhang} Z. B. Zhang, J. Y. Ying, and M. S. Dresselhaus, J. Mater. Res., \textbf{13} 1745(1998).
\bibitem {shah} H. A. Shah, M. N. S. Qureshi and N. Tsintsadze, Phys. Plasmas, \textbf{17} 032312(2010).
\bibitem {kittel} Charles Kittel, "Introduction to Solid State Physics", 8th Ed., USA: John Wiely \& Sons Corp., 1991.
\bibitem {akbarinew} M. Akbari-Moghanjoughi, IEEE Trans. Plasma Sci., \textbf{40}, 1330(2012), \\Doi: 10.1109/TPS.2012.2189137.

\end{thebibliography}
\end{document}